# Robust Graph Neural Networks for Stability Analysis in Dynamic Networks

Xin Zhang
Independent Researcher
Seattle, USA

Zhen Xu
Independent Researcher
Shanghai, China

Yue Liu
Fordham University
New York, USA

Mengfang Sun
Stevens Institute of Technological
Hoboken, USA

Tong Zhou
Northeastern University,
San Jose, USA

Wenying Sun *
Southern Methodist University
Dallas, USA

*Abstract*—In the current context of accelerated globalization and digitalization, the complexity and uncertainty of financial markets are increasing, and the identification and prevention of economic risks have become a key link in maintaining the stability of the financial system. Traditional risk identification methods often have limitations because they are difficult to cope with the multi-level and dynamically changing complex relationships in financial networks. With the rapid development of financial technology, graph neural network (GNN) technology, as an emerging deep learning method, has gradually shown great potential in the field of financial risk management. GNN can map transaction behaviors, financial institutions, individuals and their interactive relationships in financial networks into graph structures, and effectively capture potential patterns and abnormal signals in financial data through embedded representation learning. Using this technology, financial institutions can extract valuable information from complex transaction networks, identify hidden dangers or abnormal behaviors that may cause systemic risks in a timely manner, optimize decision-making processes, and improve the accuracy of risk warnings. This paper explores the economic risk identification algorithm based on the GNN algorithm, aiming to provide financial institutions and regulators with more intelligent technical tools to help maintain the security and stability of the financial market. Improving the efficiency of economic risk identification through innovative technical means is expected to further enhance the risk resistance of the financial system and lay the foundation for building a robust global financial system.

## I. INTRODUCTION

In today's digital age, the integration of digitalization into financial markets has further amplified their complexity by introducing large-scale, real-time data flows and a multitude of digital financial products. Digital platforms facilitate instantaneous transactions across borders, creating interdependencies that traditional models cannot fully capture. This rapid exchange of financial data and the increased connectivity between markets heighten the risk of cascading failures, where a disruption in one part of the network can spread quickly, affecting the global financial system. Economic risks include market volatility, credit risk, liquidity risk, cyber financial crimes, false information dissemination, and fraud. With the development of financial technology and the rise of Internet finance, it is increasingly important to identify and prevent these economic risks. Especially in financial networks, behaviors such as false transactions, financial fraud, and market manipulation pose a severe challenge to the stability of the economic system [1].

In order to address this problem, this project is committed to designing and implementing an economic risk identification algorithm based on graph neural networks (GNNs) [2]. Neural networks have been widely applied in the field of artificial intelligence, such as in image recognition [3-6]. The algorithm will transform the transaction behaviors and participants in the financial network into a graph structure and use the embedded representation learned by GNN to construct an economic risk detection model. In this way, the algorithm can extract valuable information from complex financial network data and identify potential economic risk events, thereby providing strong support for financial institutions and regulatory authorities [7].

The security of financial networks is of great significance to maintaining market stability, protecting the interests of investors, and promoting the healthy development of the economy. With the advancement of financial technology, the means of economic crimes and fraudulent behaviors are also constantly upgrading, such as financial fraud, false information dissemination, and market manipulation [8]. These behaviors not only damage the fairness of the financial market but may also have a negative impact on economic development. Therefore, the development of an efficient economic risk identification algorithm is of great practical significance for the timely detection and prevention of these bad behaviors, maintaining the normal order of the market, and ensuring economic security.

As an advanced deep learning technology, graph neural network (GNN) has demonstrated strong capabilities in processing graph-structured data [9]. GNN can capture the complex relationships between nodes and the global structural information of the graph, which provides a new perspective for understanding and analyzing economic behavior patterns in financial networks [10]. Through GNN technology, we can more accurately model and analyze economic behaviors in financial networks, thereby identifying potential risk events. This can not only improve the accuracy of economic risk identification but also prevent and reduce the occurrence of financial crimes to a certain extent, which has far-reaching significance for building a safer and more stable financial environment.

In addition, the research and development of economic risk identification algorithms based on GNN will also promote innovation and development in related technical fields. The application of GNN technology is not limited to financial risk analysis, but can also be extended to other fields, such as network security, traffic network optimization, etc. Through interdisciplinary research and cooperation, the application scope of GNN can be further expanded to promote the progress and innovation of artificial intelligence technology. This research will also promote the improvement of relevant laws and regulations, provide scientific basis and technical support for the governance of the financial market, and have important strategic significance for improving the country's economic security protection capabilities and governance level.

## II. RELATED WORK

Graph Neural Networks (GNNs) have proven to be effective tools in financial risk detection due to their ability to model the complex relationships inherent in financial networks. In this financial context, GNN represents entities such as financial institutions, individuals, or transactions as nodes in a graph, while edges denote the relationships or interactions between these entities. For example, a bank and its clients would form a node-edge structure, with transactions acting as edges. The most important relationships for analysis include capital flows between institutions, inter-company lending, and transactional patterns that may indicate fraudulent activity. By capturing these relationships, GNN is capable of identifying patterns that traditional models might overlook, such as hidden clusters of suspicious transactions. Recent studies have explored the capabilities of GNNs and deep learning in various financial risk applications, particularly credit risk assessment, fraud detection, and financial transaction analysis.

The potential of GNNs in credit risk assessment has been highlighted, demonstrating their capacity to model intricate relationships in SME transaction networks. This approach enhances the accuracy of credit risk evaluation by capturing nuanced financial interactions, making it a valuable tool for risk management [11]. In the area of fraud detection, GNNs have been integrated with reinforcement learning to create dynamic systems capable of tracking changes in user behavior and transaction patterns over time. This method is particularly useful for identifying fraudulent activities in real-time, offering a dynamic solution for financial crime prevention [12]. Adversarial learning techniques combined with GNNs have been applied to cross-domain credit risk assessment, effectively handling domain adaptation challenges and improving the robustness of credit risk models. This method strengthens the adaptability of GNNs in diverse financial sectors, reinforcing their versatility in risk prediction [13].

An adaptive feature interaction model has also been developed for credit risk prediction in digital finance, leveraging deep learning to enhance the precision of risk assessments. This work aligns with the goals of GNN-based financial data processing, where advanced feature extraction techniques are critical for accurate decision-making [14]. Spatiotemporal feature representation and multidimensional time series analysis have been applied to financial data, demonstrating how GNN-based models can capture both spatial and temporal patterns to improve risk detection. These techniques contribute to the understanding of complex financial networks and support more accurate economic risk predictions [15]. Further work has focused on optimizing the classification of financial news using Bi-LSTM and attention mechanisms, which can be incorporated into GNN-based models for detecting early signs of economic risks. This approach enhances the ability to process large-scale financial data, supporting more timely and accurate anomaly detection [16].

Graph-based models have also been applied to recommendation systems through knowledge graph attention-assisted networks, providing valuable techniques for capturing key interactions between financial entities. These approaches can be adapted to financial risk detection where understanding relationships between entities is critical [17]. The exploration of contrastive learning for knowledge-based question generation offers insights that can be integrated into financial risk models, potentially improving the quality of risk-related data analysis and enhancing the identification of anomalies in financial systems [18]. Neural networks have been successfully employed in handling heterogeneous data across multiple domains, offering advanced techniques for processing complex datasets like those found in financial networks. This work underscores the potential of deep learning in economic risk detection, where data complexity is a key challenge [19]. Finally, reducing bias in deep learning optimization has been addressed, leading to more reliable and fair GNN models. This improvement is essential for ensuring accurate predictions in financial risk assessments, particularly in markets where biases can negatively affect outcomes [20].

Collectively, these works illustrate the significant progress made in utilizing GNNs and deep learning techniques for economic risk detection. As financial systems become more complex, GNN-based models are increasingly recognized for their capacity to capture dynamic relationships, detect anomalies, and improve the accuracy of risk predictions in financial markets.

## III. METHOD

Graph Neural Network (GNN) is a deep learning model that specializes in processing graph-structured data GNN offers several advantages over traditional machine learning methods like decision trees, support vector machines (SVM), and even other neural networks. While models like decision trees and SVMs focus on isolated features, GNN captures the interconnectedness of financial entities, uncovering hidden patterns of risk spread across the network. Additionally, unlike traditional neural networks that operate on Euclidean data,

GNN excels in processing graph-structured data, making it ideal for identifying abnormal behaviors in financial networks where relationships are often non-linear and interdependent. In financial networks, the relationship between transactions and financial entities can be naturally represented as a graph, where financial entities are nodes and transactions or interactions constitute edges. Therefore, GNN has unique advantages in financial risk analysis, especially anomaly detection. GNN understands the structure of financial networks by learning embedded representations of nodes. These embedded representations can capture the local neighborhood and global network characteristics of nodes, thereby revealing patterns of financial behavior. In economic risk detection, GNN can identify nodes that are inconsistent with normal economic behavior patterns, that is, potential risk events. For example, an abnormal behavior may manifest as an unusual trading pattern or a large number of interactions with high-risk accounts. The core of GNN is the graph convolution operation, which iteratively aggregates information from neighboring nodes to learn a node's embedding. For the financial context, this allows the model to capture the influence of an entity's neighbors, such as counterparties in transactions, on its behavior. Moreover, we optimized the GCN by incorporating a multi-head attention mechanism to weigh the importance of different neighbors. This modification improves the model's ability to focus on high-risk transactions or entities that might have a disproportionate influence on the financial network's stability. The general architecture of the graph neural network is shown in Figure 1.

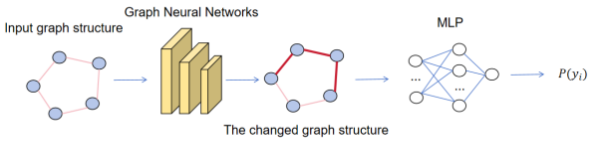

Figure 1 General architecture of graph neural network

GNN learns the embedded representation of nodes to understand the complex relationships and dependencies between nodes in the graph structure, thereby effectively identifying potential economic risk events. In GNN, each node $v_i$ in the graph represents a financial entity, while the edge $(v_i, v_j)$ in the graph represents the transaction or interaction between these entities. In order to process these graph-structured data, GNN adopts graph convolution operation, which aims to update the embedding representation of the node by aggregating the information of the node and its neighbors. The core formula of graph convolution can be expressed as:

$$h_i^{(l+1)} = \sigma(W^{(l)} \cdot AGG(\{h_j^{(l)} \mid \forall j \in N(i)\}))$$

Among them, $h_j^{(l)}$ is the embedded representation of node $v_i$ at the lth layer, $N(i)$ represents the set of neighbor nodes of node $v_i$, $W^{(l)}$ is the learnable weight matrix of the lth layer, AGG is the aggregation function of neighbor node information, and $\sigma$ is the activation function. By aggregating and updating the weight of the node and its neighbors, the embedded representation of the node can gradually reflect its position and structure in the graph.

In the application of economic risk detection, the goal of GNN is to capture abnormal behavior patterns in financial networks through this iterative update process. After multiple layers of graph convolution operations, the embedded representation $h_j^{(l)}$ of the node will be able to fully capture its role and structural characteristics in the entire financial network. These embedded representations can be used for further anomaly detection algorithms. Specifically, anomaly detection can be done by calculating the distance between the embedding representation of the node and the normal behavior pattern. Set a threshold $\tau$. If the distance between the embedding representation of the node $v_i$ and the normal behavior exceeds the threshold, the node is marked as a potential risk node. The anomaly metric can be calculated using the following formula:

$$Score ==\| h_i^{(L)} - h' \|$$

Among them, $h'$ is the mean of all normal node embedding representations. In this way, GNN can identify nodes with abnormal economic behavior patterns, thus providing strong support for financial risk management.

IV. EXPERIMENT

A. Datasets

The dataset used in this paper is an economic network structure consisting of 10,000 nodes and 32,019 edges. Each node represents a financial entity (such as a company, institution, or individual), and each edge represents a transaction or economic relationship between these entities. The transaction relationships in the dataset are derived from public financial datasets, reflecting the capital flow, investment cooperation, or debt relationship of financial entities in the market. Initially, we employed the linked data approach to handle the dataset, dismantle data silos, and facilitate a more integrated and thorough method of analysis [21]. By analyzing the complex structure of these nodes and edges, it is possible to reveal normal economic activity patterns and help identify economic behaviors or abnormal trading activities that may involve high risks. In the context of economic risk detection, this graph structure can reveal normal cooperation patterns between financial entities while identifying high-risk entities that exhibit atypical behaviors. Some financial entities may exhibit abnormal connection patterns, such as abnormally frequent capital flows, sudden establishment or disconnection of financial connections with a large number of other entities, or the formation of unusual substructures in the economic network, such as closely connected risk groups. These behaviors may indicate potential financial crimes, money laundering, or fraudulent activities. With 10,000 nodes, the dataset can be regarded as a medium-sized economic network, which is large enough to reveal the complexity of financial transactions, while the data size is moderate and easy to process and model. The existence of 32,019 edges indicates that the network contains rich economic interaction information, which provides sufficient basic data for the application of graph analysis technology.

*B. Experimental Detail*

The experimental setup is mainly centered around the economic risk identification algorithm based on a graph neural network (GNN) [22]. First, we defined a set of hyperparameters through the argparse library, including the number of financial entities entity_num, the embedding dimension embed_size, the number of attention heads attention_heads, the dimension of each attention head head_dim, the number of hyperedges of transaction relations hypernum, and the dropout ratio of the embedding layer. These parameters provide the necessary configuration for the training of the algorithm and allow users to adjust them according to the specific needs of the economic network. In the experiment, we selected an economic transaction network dataset containing 10,000 financial entities. The embedding dimension is set to 128, the number of attention heads is 8, the dimension of each head is 50, and the dropout ratio is set to 0.2. In the model training stage, the Adam optimizer is used, which is an optimization algorithm with high computational efficiency and suitable for large-scale datasets, especially in economic data scenarios. It can converge quickly. The learning rate lr is set to 0.00003, the training cycle is 20 rounds, and each round of data is divided into 32 batches for training. In order to improve the training efficiency and the generalization ability of the model, we define a custom MyDataset class to create a PyTorch DataLoader object, which is responsible for batch loading and random shuffling of data so that economic transaction data can be better used for risk identification. The loss function uses the cross-entropy loss function CrossEntropyLoss to calculate the difference between the economic risk label predicted by the model and the true risk label. After the training is completed, the model parameters and training loss are saved for further analysis and evaluation. In the testing phase, the performance of the model of the test set is evaluated, and the accuracy on the test set is used as an important indicator to measure the model's ability to identify economic risks.

*C. Experimental Result*

Before evaluating the performance of the model, we first give a training process diagram of the model loss decreasing with epoch.

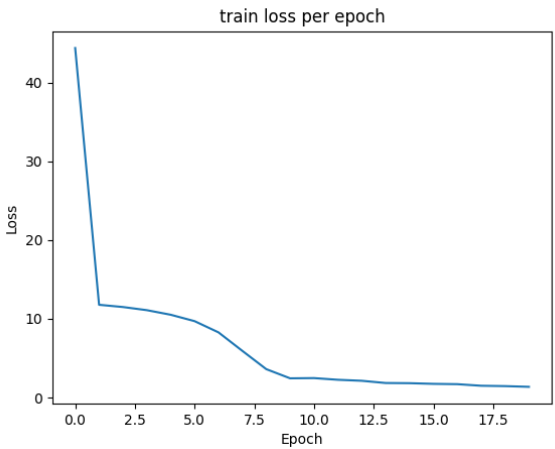

Figure 2 Loss function decline process diagram

The experimental analysis focuses on evaluating the performance of the economic risk identification algorithm based on graph neural network (GNN) during the training process. According to the curve of the decline of the loss function, we can clearly see that with the increase of training rounds (epochs), the loss value of the model gradually decreases. This trend shows that the model is constantly optimizing and can effectively learn the complex relationships in the economic network. The higher loss value in the initial stage indicates that the model is in the early stage of learning the economic risk pattern, but as the training progresses, the model gradually adapts to the characteristics of the data set and the loss value decreases rapidly. This reflects the efficiency of the optimization algorithm and the adaptability of the model structure in identifying economic risk patterns.

As the training progresses, the rate of decline of the loss value gradually slows down, indicating that the model is close to its best performance. In the final stage of training, the loss value tends to stabilize and forms a nearly horizontal curve, which is a typical learning curve, indicating that the model has achieved a good fit and the weight parameters are adjusted to an ideal state. In the context of economic risk identification, the continuous decline and final stabilization of the loss function indicate that the GNN model successfully learns the differences between normal behavior and potential risk behavior in the economic network and effectively transforms these differences into prediction results. This performance improvement is crucial for identifying and responding to potential economic risks in advance, and helps build a more stable and secure financial environment. Overall, the decrease in the loss function provides positive feedback for model training, proving the success of the model structure design and the effectiveness of the optimization strategy. This result strengthens our confidence in the GNN model in the task of economic risk identification and provides a solid foundation for future model improvement and optimization.

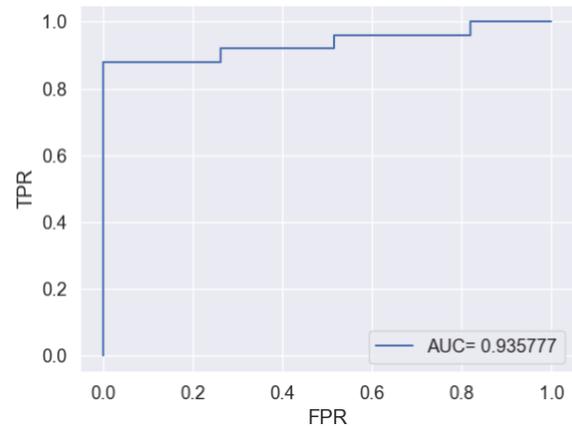

Figure 3 ROC indicator analysis

As can be seen from the figure, the AUC score of the model is 0.935777. This high score indicates that the model performs well in distinguishing economic risks from normal economic behaviors. As can be seen from the figure, the ROC curve is close to the upper right corner of the coordinate axis, indicating that the model has a high true positive rate (True Positive Rate, TPR) and a low false positive rate (False Positive Rate, FPR).

When FPR is 0, TPR has begun to increase, which means that the model can correctly identify some real economic risk behaviors without generating false positives. As FPR increases, TPR continues to rise, showing that the model maintains a high recognition ability at different thresholds. AUC values close to 1.0 indicate that the classification performance of the model is very good. In the task of economic risk identification, this means that the model can identify potential economic risks with a very high probability, while minimizing the probability of misjudging normal economic behaviors as risks. This efficient risk identification ability is of great significance for timely warning of potential crises in economic activities, and helps decision-makers take timely response measures to reduce economic losses.

In addition, the high AUC value also indicates that the model has good generalization ability and can adapt to economic activities and risk patterns under different market conditions. For the characteristics of rapid changes and complex interactions in the economic environment, the GNN model extracts features that help identify economic risks by learning the relationships and network structures between various economic entities, and performs well on the test set. In summary, the excellent performance of the GNN-based economic risk identification algorithm in the AUC indicator verifies its potential as an efficient and reliable economic risk detection tool. This result not only proves the effectiveness of the model design, but also provides solid support for the future deployment of applications on larger and more diverse economic data sets.

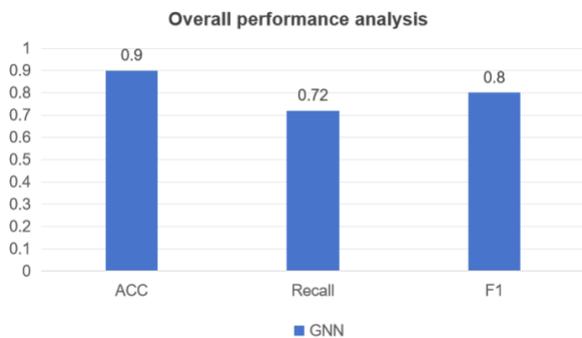

Figure 4 Overall performance analysis chart

The experimental performance analysis focuses on the overall indicators of the economic risk identification system based on a graph neural network (GNN), including precision, recall, and F1 score. According to the provided charts, the GNN model performs well on these key performance indicators, demonstrating its efficiency and reliability in the task of economic risk detection. Accuracy reflects the proportion of economic risks and normal economic behaviors correctly identified by the model. The high accuracy of the GNN model shows that it can effectively distinguish between potential economic risks and normal economic activities. This is crucial in practical economic applications because it ensures that most economic behaviors are correctly classified and avoids economic decision-making errors that may be caused by misjudgment. Recall (also known as the true positive rate) measures the proportion of actual risk behaviors identified by the model among all real risk behaviors. The high recall of the GNN model shows that it can accurately capture most of the real economic risk behaviors, which is of great significance for the early detection of potential economic crises and timely response measures. The F1 score is the harmonic mean of precision and recall, providing a comprehensive evaluation of the model's accuracy and coverage. The F1 score of the GNN model also performs well, indicating that the model maintains a high degree of comprehensiveness in identifying economic risks while efficiently identifying economic risks. A high F1 score means that the model not only performs well in accurately identifying risks, but also effectively covers most potential risk behaviors in a variety of economic environments.

In the context of economic risk identification, the GNN model can accurately extract features that help identify risks by learning complex interaction patterns between economic entities and graph structure information. This provides strong technical support for economic risk management and early warning systems, and helps to protect the stability and security of the economic system. In summary, the GNN model performs well in accuracy, recall, and F1 scores, fully demonstrating its potential for application in economic risk identification. The high values of these performance indicators not only verify the model's predictive ability, but also provide a solid foundation for future applications in more complex economic environments. These results show that the GNN model is a reliable tool for economic risk identification that can provide critical support to decision-makers.

V. CONCLUSION

By building an economic risk identification algorithm based on graph neural networks, we are not only able to effectively capture most real economic risk behaviors, but also significantly improve the accuracy and timeliness of risk warnings. The development of a graph neural network (GNN)-based economic risk identification algorithm is crucial for enhancing the resilience of financial systems, particularly in the context of the United States, where the stability of financial markets is integral to both national and global economic health. Given the complexity and interconnectedness of U.S. financial networks, the ability to identify risks—such as market manipulation, financial fraud, and systemic vulnerabilities—has far-reaching implications for safeguarding the economy. The algorithm's excellent performance in key indicators such as accuracy, recall, and F1 score fully proves its effectiveness and reliability in a variety of economic environments. In addition, the performance of high AUC values further demonstrates that the algorithm has strong generalization ability, can flexibly adapt to different market conditions, and accurately identifies potential risk patterns in economic activities. This result verifies the great potential of the GNN algorithm in economic risk detection and shows that it can play an important role in complex and ever-changing financial markets. More importantly, this lays a solid foundation for future applications on larger-scale and more diverse economic data sets, providing valuable experience and direction for further optimizing and expanding risk detection tools.